\documentclass[aps,prd,preprintnumbers,showpacs,nofootinbib,onecolumn,showkeys]{revtex4-1}

\usepackage{amsmath}

\begin{document}
\title{Extended Reissner-Nordstr\"{o}m solutions sourced by dynamical torsion}

\author{Jose A. R. Cembranos$^{1,2,}$\footnote{E-mail: cembra@fis.ucm.es},
and Jorge Gigante Valcarcel$^{1,}$\footnote{E-mail: jorgegigante@ucm.es}
}

\affiliation{$^{1}$Departamento de  F\'{\i}sica Te\'orica I and UPARCOS, Universidad Complutense de Madrid, E-28040 Madrid, Spain\\
$^{2}$Departamento de  F\'{\i}sica, Universidade de Lisboa,
%Edif\'{\i}cio C8, Campo Grande,
P-1749-016 Lisbon, Portugal
}

%\date{\today}


% 04.70.Bw Classical black holes
% 04.40.-b Self-gravitating systems; continuous media and classical fields in curved spacetime
% 04.50.Kd	Modified theories of gravity
% 04.50.-h 	Higher-dimensional gravity and other theories of gravity (see also 11.25.Mj Compactification and four-dimensional models, 11.25.Uv D branes)
% 04.20.Jb        Exact solutions
% 11.10.Lm Nonlinear or nonlocal theories and models

% 03.50.De Classical electromagnetism, Maxwell equations
% 05.45.Yv Solitons
% \pacs{04.50.Kd, 04.50.-h, 98.80.-k}
%
%04.25.Nx 	Post-Newtonian approximation; perturbation theory; related approximations
% 95.36.+x 	Dark energy (see also 98.80.-k Cosmology)
% 98.80.-k	 Cosmology (see also section 04 General relativity and gravitation; for origin and evolution of galaxies, see 98.62.Ai; for elementary particle and nuclear processes, see 95.30.Cq; for dark matter, see 95.35.+d; for dark energy, see 95.36.+x; for superclusters and large-scale structure of the Universe, see 98.65.Dx)
%
%
%04.20.-q	         Classical general relativity (see also 02.40.-k Geometry, differential geometry, and topology)
%04.20.Cv	Fundamental problems and general formalism
% 98.80.Jk        Mathematical and relativistic aspects of cosmology
% 04.40.Dg Relativistic stars: structure, stability, and oscillations
% 11.27.+d Extended classical solutions; cosmic strings, domain walls, texture.
% 11.25.-w: Strings and branes
%%%%%%%%%%%%%

\begin{abstract}

We find a new exact vacuum solution in the framework of the Poincar\'e Gauge field theory with massive torsion. In this model, torsion operates as an independent field and introduces corrections to the vacuum structure present in General Relativity. The new static and spherically symmetric configuration shows a Reissner-Nordstr\"{o}m-like geometry characterized by a spin charge. It extends the known massless torsion solution to the massive case. The corresponding Reissner-Nordstr\"{o}m-de Sitter solution is also compatible with a cosmological constant and additional $U(1)$ gauge fields.

\end{abstract}


\maketitle

\section{Introduction}

The fundamental relation of the energy and momentum of matter with the space-time geometry is one of the most important foundations of General Relativity (GR). Namely, the energy-momentum tensor acts as the source of gravity, which is appropriately described in terms of the curvature tensor. In an analogous way, it may be expected that the intrinsic angular momentum of matter may also act as an additional source of the interaction and extend such a geometrical scheme. 

Poincar\'e Gauge (PG) theory of gravity is the most consistent extension of GR that provides a suitable correspondence between spin and the space-time geometry by assuming an asymmetric affine connection defined within a Riemann-Cartan (RC) manifold (i.e. endowed with curvature and torsion) \cite{Hehl,Blag-Hehl}. It represents a gauge approach to gravity based on the semidirect product of the Lorentz group and the space-time translations, in analogy to the unitary irreducible representations of relativistic particles labeled by their spin and mass, respectively. Then not only an energy-momentum tensor of matter arises from this approach, but also a non-trivial spin density tensor that operates as source of torsion and allows the existence of a gravitating antisymmetric component of the former, which may induce changes in the geometrical structure of the space-time, as the rest of the components of the mentioned tensor. This fact contrasts with the established by GR, where all the possible geometrical effects occurred in the Universe can be only provided by a symmetric component of the energy-momentum tensor, despite the existence of dynamical configurations endowed with asymmetric energy-momentum tensors \cite{Papapetrou:1949,Sciama}.

Accordingly, a gauge invariant Lagrangian can be constructed from the field strength tensors to introduce the extended dynamical effects of the gravitational field. In this sense, it is well-known that the role of torsion depends on the order of the mentioned field strength tensors present in the Lagrangian, in a form that only quadratic or higher order corrections in the curvature tensor involve the presence of a non-trivial dynamical torsion, whose effects can propagate even in a vacuum space-time.

Likewise, the distinct restrictions on the Lagrangian parameters lead to a large class of gravitational models where an extensive number of particular and fundamental differences may arise. For example, in analogy to the standard approach of GR, it was shown that the Birkhoff's theorem is satisfied only in certain cases of the PG theory \cite{Nev,Rauch-Nieh}. Indeed, the dynamical role of the new degrees of freedom involved in such a theory can modify the space-time geometry and even predominate in their respective domains of applicability. The search and study of exact solutions are therefore essential in order to improve the understanding and physical interpretation of the new framework.

A large class of exact solutions have been found since the formulation of the theory, especially for the case of static and spherically symmetric vacuum space-times, where one of the most primary and remarkable solutions is the so called Baekler solution, associated with a sort of PG models that encompass a weak-field limit with an additional confinement type of potential besides the Newtonian one \cite{Held}, giving rise to a Schwarzschild-de Sitter geometry in analogy to the effect caused by the presence of a cosmological constant in the regular gravity action \cite{Baekler}. Furthermore, additional results have also been systematically obtained for a large class of PG configurations, such as axisymmetric space-times, cosmological systems or generalized gravitational waves (see \cite{Blag-Hehl,Obukhov,Obukhovv,Blagojevic:2017wzf} and references therein).

Recently, the authors of this work found a new exact solution with massless torsion associated with a PG model containing higher order corrections quadratic in the curvature tensor, in such a way that the standard framework of GR is naturally recovered when the total curvature satisfies the first Bianchi identity of the latter. This construction ensures that all the new propagating degrees of freedom introduced by the model fall on the torsion field, so that this quantity extend the domain of applicability of the standard case. Thus, it was shown that the regular Schwarzschild geometry provided by the Birkhoff's theorem of GR can be replaced by a Reissner-Nordstr\"{o}m (RN) space-time with RC Coulomb-like curvature when this sort of dynamical torsion is considered \cite{CembGig}. This result contrasts with other post-Riemannian solutions, such as the derived in the framework of the Metric-Affine Gauge (MAG) theory, where the non-metricity tensor can involve an analogous vacuum RN configuration \cite{Tresguerres,Hehl-Mac}. In addition, it is reasonable to expect that such a configuration may be extended for the case where additional non-vanishing mass modes of the torsion tensor are present in the Lagrangian, in order to analyze the equivalent PG model with massive torsion. As we will show, we have found the associated RN solution with massive torsion and generalized the previous approach according to the scheme performed in that simpler case.

This paper is organized as follows. First, in Section II, we introduce our PG model with massive torsion and briefly describe its general mathematical foundations. The analysis and application of the resulting field equations in the static spherically symmetric space-time is shown in Section III, in order to find the appropriate vacuum solutions for the selected case. In section IV, we present the required new PG solution with massive torsion and extend our previous results related to the massless case. We present the conclusions of our work in Section V. Finally, we detail in Appendix A the geometrical quantities involved in the vacuum field equations associated with this model.

Before proceeding to the main discussion and general results, we briefly introduce the notation and physical units to be used throughout this article. Latin $a,b$ and greek $\mu, \nu$ indices refer to anholonomic and coordinate basis, respectively. We use notation with tilde for magnitudes including torsion and without tilde for torsion-free quantities. On the other hand, we will denote as $P_{a}$ the generators of the space-time translations as well as $J_{a b}$ the generators of the space-time rotations and assume their following commutative relations:

\begin{equation}
\left[P_{a},P_{b}\right]=0\,,
\end{equation}

\begin{equation}
\left[P_{a},J_{bc}\right]=i\,\eta_{a[b}\,P_{c]}\,,
\end{equation}

\begin{equation}
\left[J_{ab},J_{cd}\right]=\frac{i}{2}\,\left(\eta_{ad}\,J_{bc}+\eta_{cb}\,J_{ad}-\eta_{db}\,J_{ac}-\eta_{ac}\,J_{bd}\right)\,.
\end{equation}
Finally, we will use Planck units ($G=c=\hbar=1)$ throughout this work.

\section{Quadratic Poincar\'e gauge gravity model with massive torsion}

We start from the general gravitational action associated with our original PG model and incorporate the three independent quadratic scalar invariants of torsion into this expression, which represent the mass terms of the mentioned quantity:

\begin{eqnarray}\label{Lagrangian}
S &=& \; \frac{1}{16 \pi}\int d^4x \sqrt{-g}
\Bigl[
\mathcal{L}_{m}-\tilde{R}-\frac{1}{4}\left(d_{1}+d_{2}\right)\tilde{R}^2-\frac{1}{4}\left(d_{1}+d_{2}+4c_{1}+2c_{2}\right)\tilde{R}_{\lambda \rho \mu \nu}\tilde{R}^{\mu \nu \lambda \rho}+c_{1}\tilde{R}_{\lambda \rho \mu \nu}\tilde{R}^{\lambda \rho \mu \nu}
\Bigr.
\nonumber\\
& &
\Bigl.
\;\;\;\;\;\;\;\;\;\;\;\;\;\;\;\;\;\;\;\;\;\;\;\;\;\;\;
+c_{2}\tilde{R}_{\lambda \rho \mu \nu}\tilde{R}^{\lambda \mu \rho \nu}
+d_{1}\tilde{R}_{\mu \nu}\tilde{R}^{\mu\nu}+d_{2}\tilde{R}_{\mu\nu}\tilde{R}^{\nu\mu}+\alpha \, T_{\lambda \mu \nu}T^{\lambda \mu \nu}+\beta \, T_{\lambda \mu \nu}T^{\mu \lambda \nu}+\gamma \, T^{\lambda}\,_{\lambda \nu}T^{\mu}\,_{\mu}\,^{\nu}
\Bigr]\,,
\label{actioneq}
\end{eqnarray}
where $c_{1},c_{2},d_{1},d_{2},\alpha,\beta$ and $\gamma$ are constant parameters.

The field strength tensors above derive from the gauge connection of the Poincar\'e group $ISO (1,3)$, which can be expressed in terms of the generators of translations and local Lorentz rotations in the following way:

\begin{equation}A_{\mu}=e^{a}\,_{\mu}P_{a}+\omega^{a b}\,_{\mu}J_{a b}\,,\end{equation}
where $e^{a}\,_{\mu}$ is the vierbein field and $\omega^{a b}\,_{\mu}$ the spin connection of a RC manifold, related to the metric tensor and the metric-compatible affine connection as usual \cite{Yepez}:

\begin{equation}
g_{\mu \nu}=e^{a}\,_{\mu}\,e^{b}\,_{\nu}\,\eta_{a b}\,,
\end{equation}

\begin{equation}
\omega^{a b}\,_{\mu}=e^{a}\,_{\lambda}\,e^{b \rho}\,\tilde{\Gamma}^{\lambda}\,_{\rho \mu}+e^{a}\,_{\lambda}\,\partial_{\mu}\,e^{b \lambda}\,.
\end{equation}

The affine connection is decomposed into the torsion-free Levi-Civita connection and a contortion component, which transforms as a tensor due to the tensorial nature of torsion since it describes the antisymmetric part of the affine connection:

\begin{equation}
\tilde{\Gamma}^{\lambda}\,_{\mu \nu} = {\Gamma}^{\lambda}\,_{\mu \nu} + K^{\lambda}\,_{\mu \nu}\,.
\end{equation}

Thus, the presence of torsion potentially introduces changes in the properties of the gravitational interaction and it involves the following $ISO (1,3)$ gauge field strength tensors:

\begin{equation}F^{a}\,_{\mu \nu}=e^{a}\,_{\lambda}\,T^{\lambda}\,_{\nu \mu}\,,\end{equation}

\begin{equation}F^{a b}\,_{\mu \nu}=e^{a}\,_{\lambda}e^{b}\,_{\rho}\,\tilde{R}^{\lambda \rho}\,_{\mu \nu}\,,\end{equation}
where $T^{\lambda}\,_{\mu \nu}$ and $\tilde{R}^{\lambda \rho}\,_{\mu \nu}$ are the components of the torsion and the curvature tensor, respectively:

\begin{equation}T^{\lambda}\,_{\mu \nu}=2\tilde{\Gamma}^{\lambda}\,_{[\mu \nu]}\,,\end{equation}

\begin{equation}\tilde{R}^{\lambda}\,_{\rho \mu \nu}=\partial_{\mu}\tilde{\Gamma}^{\lambda}\,_{\rho \nu}-\partial_{\nu}\tilde{\Gamma}^{\lambda}\,_{\rho \mu}+\tilde{\Gamma}^{\lambda}\,_{\sigma \mu}\tilde{\Gamma}^{\sigma}\,_{\rho \nu}-\tilde{\Gamma}^{\lambda}\,_{\sigma \nu}\tilde{\Gamma}^{\sigma}\,_{\rho \mu}\,.\end{equation}

Therefore, within this framework, torsion appears naturally related to the translations whereas curvature is related to the rotations, as expected. Furthermore, both quantities can decompose into distinct modes by computing their irreducible representations under the Lorentz group \cite{Gambini,Hay-Shi}. Specifically, torsion can be divided into three irreducible components: a trace vector $T_{\mu}$, an axial vector $S_{\mu}$ and a traceless and also pseudotraceless tensor $q^{\lambda}\,_{\mu \nu}$:

\begin{equation}
T^{\lambda}\,_{\mu \nu}=\frac{1}{3}\left(\delta^{\lambda}\,_{\nu}T_{\mu}-\delta^{\lambda}\,_{\mu}T_{\nu}\right)+\frac{1}{6}\,g^{\lambda \rho}\varepsilon\,_{\rho \sigma \mu \nu}S^{\sigma}+q^{\lambda}\,_{\mu \nu}\,,
\end{equation}
where $\varepsilon\,_{\rho \sigma \mu \nu}$ is the four-dimensional Levi-Civita symbol.

Hence, each of the cited modes can be massive or massless, what can be implemented in the general action of the theory by introducing the corresponding explicit torsion square pieces, as it is shown in the Expression (\ref{actioneq}). Then, the extended field equations can be derived by performing variations with respect to the gauge potentials, as usual. In addition, the resulting system of equations can be simplified without loss of generality by the Gauss-Bonnet theorem in RC spaces \cite{Nieh,Hayashi}. Namely, the following combination quadratic in the curvature tensor acts as a total derivative of a certain vector $V^\mu$ in the previous gravitational action:

\begin{equation}\sqrt{- g}\,\left(\tilde{R}^{2}+\tilde{R}_{\lambda \rho \mu \nu}\tilde{R}^{\mu \nu \lambda \rho}-4\tilde{R}_{\mu \nu}\tilde{R}^{\nu \mu}\right)=\partial_\mu V^\mu\,.\end{equation}

Thereby, this constraint allows to reduce the gravitational action and to obtain the following system of variational equations:

\begin{equation}
X1_{\mu}\,^{\nu}+16 \pi \theta_{\mu}\,^{\nu}= 0\,,\end{equation}

\begin{equation}
X2_{[\mu \lambda]}\,^{\nu}+16\pi S_{\lambda \mu}\,^{\nu}= 0\,,\end{equation}
where $X1_{\mu}\,^{\nu}$ and $X2_{[\mu \lambda]}\,^{\nu}$ are tensorial functions depending on the RC curvature and the torsion tensor, which are defined in Appendix A, whereas $\theta_{\mu}\,^{\nu}$ and $S_{\lambda \mu}\,^{\nu}$ are the canonical energy-momentum tensor and the spin density tensor, respectively:

\begin{equation}\theta_{\mu}\,^{\nu}=\frac{e^{a}\,_{\mu}}{16 \pi \sqrt{- g}}\frac{\delta\left(\mathcal{L}_{m}\sqrt{- g}\right)}{\delta e^{a}\,_{\nu}}\,,\end{equation}

\begin{equation}S_{\lambda \mu}\,^{\nu}=\frac{e^{a}\,_{\lambda}e^{b}\,_{\mu}}{16 \pi \sqrt{- g}}\frac{\delta\left(\mathcal{L}_{m}\sqrt{- g}\right)}{\delta \omega^{a b}\,_{\nu}}\,.\end{equation}

These quantities act as sources of gravity and constitute the natural generalization of the conserved Noether currents associated with the external translations and rotations of the Poincar\'e group in a Minkowski space-time \cite{Kosmann-Schwarzbach}. Indeed, it is straightforward to note from the field equations above the fulfillment of the following conservation laws:

\begin{equation}
\nabla_{\nu}\theta_{\mu}\,^{\nu}+K_{\lambda \rho \mu}\theta^{\rho \lambda}+\tilde{R}_{\lambda \rho \nu \mu}\,S^{\lambda \rho \nu}=0\,,
\end{equation}

\begin{equation}
\nabla_{\mu}S_{\lambda \rho}\,^{\mu}+2K^{\sigma}\,_{[\lambda | \mu}S_{|\rho] \sigma}\,^{\mu}-\theta_{[\lambda \rho]}=0\,.
\end{equation}

Therefore, the canonical energy-momentum tensor generally
contains an antisymmetric component even when the notions of curvature and torsion are neglected (i.e. in the framework of Special Relativity):

\begin{equation}
\partial_{\nu}\theta_{\mu}\,^{\nu}=0\,,
\end{equation}

\begin{equation}
\partial_{\mu}M_{\lambda \rho}\,^{\mu}+\partial_{\mu}S_{\lambda \rho}\,^{\mu}=0\,,
\end{equation}
where $M_{\lambda \rho}\,^{\mu}=x_{[\lambda}\,\theta_{\rho]}\,^{\mu}$ is the orbital angular momentum density, whose divergence is trivially proportional to the mentioned antisymmetric part of the canonical energy-momentum tensor:

\begin{equation}
\partial_{\mu}M_{\lambda \rho}\,^{\mu}=\theta_{[\rho \lambda]}\,.
\end{equation}

Thus, as it is shown, there exists a complete correspondence between the main currents of matter sources and the space-time geometry in the framework of PG theory. However, the theoretical construction present in GR encodes all the possible geometrical effects, derived by the presence of the gravitational field, only into the symmetric part of the canonical energy-momentum tensor of matter. Specifically, it postulates the symmetrized Belinfante-Rosenfeld energy-momentum tensor as the unique material quantity coupled to gravity \cite{Rosenfeld}:

\begin{equation}
T_{\mu \nu}=\theta_{\mu \nu}-\nabla_{\lambda}S_{\mu \nu}\,^{\lambda}-\nabla_{\lambda}S^{\lambda}\,_{\mu \nu}-\nabla_{\lambda}S^{\lambda}\,_{\nu \mu}\,,
\end{equation}
and omits from the gravitational scheme all the possible dynamical contributions provided by the rest of features of matter. Some remarkable implications derived by this post-Riemannian approach involve the prevention of space-time singularities and the generation of an accelerating cosmological expansion in terms of the torsion field, among others \cite{Poplawski,Magueijo,Chen:2009at,Minkevich,Lu}. In this sense, apart from its potential influence in the cosmological and astrophysical arena, the space-time torsion represents a fundamental quantity that may improve our understanding on the correspondence between geometry and physics, what it means that any kind of dynamical aspect associated with it may be crucial to identify its different roles or to detect it.

Concerning the vacuum structure of the theory, the material tensors above vanish and it is sufficient to deal with the following system of equations:

\begin{equation}X1_{\mu}\,^{\nu} = 0\,,\end{equation}

\begin{equation}\label{connectioneqq}X2_{[\mu \lambda]}\,^{\nu} = 0\,.\end{equation}

It is straightforward to note that the standard approach of GR is completely recovered when the first Bianchi identity of such a theory is fulfilled by the total curvature (i.e. $\tilde{R}^{\lambda}\,_{[\mu \nu \rho]}=0$) and all the mass coefficients of torsion vanish. However, in the massless torsion solution \cite{CembGig}, it was shown that such a limit can be obtained by switching off the dynamical axial component of the torsion tensor, so that even for the case where both the trace vector and the tensorial component of torsion are massless, the same procedure may be trivially applied in presence of a massive axial component of torsion.

\section{Space-time symmetries and consistency constraints}

In order to solve the vacuum field equations of the theory for a static and spherically symmetric space-time, we consider the corresponding line element and tetrad basis as follows:

\begin{equation}ds^2=\Psi_{1}(r)\,dt^2-\frac{dr^2}{\Psi_{2}(r)}-r^2\left(d\theta_{1}^{2}+\sin^{2}{\theta_{1}}d\theta_{2}^{2}\right)\,,\end{equation}

\begin{equation}
e^{\hat{t}}=\sqrt{\Psi_{1}(r)}\,dt\,,\;\;\;
e^{\hat{r}}=\frac{dr}{\sqrt{\Psi_{2}(r)}}\,,\;\;\;
e^{\hat{\theta_{1}}}=r\,d \theta_{1}\,,\;\;\;
e^{\hat{\theta_{2}}}=r\sin\theta_{1} \, d \theta_{2}\,;
\end{equation}
with $0 \leq \theta_{1} \leq \pi$ and $0 \leq \theta_{2} \leq 2\pi$.

The intrinsic relations between curvature and torsion involve that the latter is also influenced by the space-time symmetries and it must satisfy the condition $\mathcal{L}_{\xi}T^{\lambda}\,_{\mu \nu}=0$ (i.e. the Lie derivative in the direction of the Killing vector $\xi$ on $T^{\lambda}\,_{\mu \nu}$ vanishes). Indeed, this constraint ensures that the covariant derivative commutes with the Lie derivative, what in turn preserves the invariance of the curvature tensor under isometries.

Therefore, the static spherically symmetric torsion acquires the following structure \cite{Rauch-Nieh,Sur-Bhatia}:

\begin{eqnarray}
T^{t}\,_{t r}&=&a(r) \;\;\, ,\nonumber\\
T^{r}\,_{t r}&=&b(r) \;\;\; ,\nonumber\\
T^{\theta_{k}}\,_{t \theta_{k}}&=&c(r) \;\;\; ,\nonumber\\
T^{\theta_{k}}\,_{r \theta_{k}}&=&g(r) \;\;\; ,\nonumber\\
T^{\theta_{k}}\,_{t \theta_{l}}&=&e^{a \theta_{k}}\,e^{b}\,_{\theta_{l}}\,\epsilon_{a b}\, d (r) \;\;\; , \nonumber\\
T^{\theta_{k}}\,_{r \theta_{l}}&=&e^{a \theta_{k}}\,e^{b}\,_{\theta_{l}}\,\epsilon_{a b}\, h (r) \;\;\; ,  \nonumber\\
T^{t}\,_{\theta_{k} \theta_{l}}&=&\epsilon_{k l} \, k (r)\,\sin\theta_{1} \;\;\; , \nonumber\\
T^{r}\,_{\theta_{k} \theta_{l}}&=&\epsilon_{k l} \, l (r)\,\sin\theta_{1} \;\;\; ;
\end{eqnarray}
where $a,b,c,d,g,h,k$ and $l$ are eight arbitrary functions depending only on r; $k,l=1,2$ with $k \neq l$ and $\epsilon_{a b}$ is the two-dimensional Levi-Civita symbol:

\begin{equation}
\epsilon_{a b} = \left\{
\begin{array}{l}
+1\,, \;\;\;\; \text{for} \;\;\; a\,b=1\,2. \\
-1\,, \;\;\;\; \text{for} \;\;\; a\,b=2\,1. \\
\;\;\;0\,, \;\;\;\; \text{for all other combinations}.
\end{array}
\right.
\end{equation}

These symmetry properties strongly reduce the possible classes of solutions, but even though the field equations constitute a highly nonlinear system involving a large number of degrees of freedom, so that the problem turns out to be still very complicated and furthermore underdetermined. In fact, one of the features associated with a large number of PG models is the existence of a high geometrical freedom, where it is possible to find solutions depending on arbitrary functions and thereby underdetermined by the variational equations \cite{Lenzen,Chen,Zhytnikov}. It is worthwhile to stress that, for the particular case given by the presence of a dynamical massless torsion, the traceless of the tetrad field equations requires the vanishing of the torsion-free scalar curvature, which in turn represents a strong geometrical constraint involving the degrees of freedom of the metric tensor alone. Furthermore, in presence of an external Coulomb electric field, the compatibility with the Maxwell field equations in spherically symmetric space-times requires the additional constraint given by $\Psi_{1} = \Psi_{2}$, so that in this case the geometry acquires the form of a RN space-time and such a type of arbitrariness does not emerge, in contrast with other PG models with explicit torsion square pieces. In this sense, as previously stressed, we simply extend our previous results with massless torsion to a generic PG model with these torsion square corrections, what it means an easy way to obtain solutions due to the analyses performed in that simpler case. On the other hand, it is worthwhile to emphasize that the existence and unicity of solutions within these torsion models can be established under appropriate energy conditions \cite{Cembranos:2016xqx}.

According to the massless torsion scheme, it is always possible to impose an additional constraint by applying the weak-field approximation for the torsion tensor through the trace of Eq. (\ref{connectioneqq}). This restriction ensures that our PG model appropriately encompasses such a limit. Then, by neglecting torsion terms of second order, the equations of motion for the torsion tensor in linear approximation read

\begin{equation}\label{linear}\nabla_{\mu}\nabla^{\mu}T^{\nu}\,_{\lambda \nu}+\nabla_{\mu}\nabla_{\nu}T^{\nu \mu}\,_{\lambda}-\nabla_{\mu}\nabla_{\lambda}T^{\nu \mu}\,_{\nu} = \frac{2\alpha+\beta+3\gamma+2}{4c_{1}+c_{2}+2d_{1}}\,T^{\nu}\,_{\lambda \nu}\,.\end{equation}

In the special case where $2\alpha+\beta+3\gamma+2=0\,$, it turns out that the mass modes of torsion do not contribute to the weak-field approximation and then this constraint reduces to the following relation among the torsion and metric components:

\begin{equation}
\label{rel}
b(r)=rc\,'(r)+c(r)+\frac{p}{r}\,
\sqrt{\frac{\Psi_{1}(r)}{\Psi_{2}(r)}}\,,
\end{equation}
where $p$ is an integration constant.

In analogy to the massless torsion case \cite{CembGig}, we demand the condition $\Psi_{1} = \Psi_{2} \equiv \Psi(r)$ to guarantee the compatibility requirement with external electric and magnetic fields, as in the standard Einstein-Maxwell framework of GR. Finally, we also require the avoidance of undesirable singularities from any solution $F^{a}\,_{b c}$ referred to the rotated basis $\vartheta^{a}=\Lambda^{a}\,_{b}e^{b}$ given by the following vector fields:

\begin{eqnarray}
\vartheta^{\hat{t}}&=&\frac{1}{2}\left\{ \left[\Psi(r)+1\right]\,dt+\left[1-\frac{1}{\Psi(r)}\right]\,dr \right\} \;\; ;\nonumber\\
\vartheta^{\hat{r}}&=&\frac{1}{2}\left\{ \left[\Psi(r)-1\right]\,dt+\left[1+\frac{1}{\Psi(r)}\right]\,dr \right\} \;\; ;\nonumber\\
\vartheta^{\hat{\theta}_{1}}&=&r\,d \theta_{1} \;\; ;\nonumber\\%
\vartheta^{\hat{\theta_{2}}}&=&r\sin\theta_{1} \, d \theta_{2} \;\;.
\end{eqnarray}

Accordingly, in order to avoid geometrical divergences in the roots of the metric function $\Psi(r)$, the following relations among the torsion components are taken into account:

\begin{equation}
\label{rel2}
b(r) = a(r)\,\Psi(r)\,,\;\;\;c(r) = - \, g(r)\,\Psi(r)\,,\;\;\;d(r) = - \, h(r)\,\Psi(r)\,,\;\;\;l(r) = k(r)\,\Psi(r)\,.
\end{equation}

It is worthwhile to note that these constraints involve the vanishing of the three independent quadratic torsion invariants. Namely, in terms of its irreducible components:

\begin{equation}
T_{\mu}T^{\mu}=S_{\mu}S^{\mu}=q_{\lambda \mu \nu}\,q^{\lambda \mu \nu}=0\,.
\end{equation}

Furthermore, the additional quartic torsion invariants also vanish under these conditions:

\begin{equation}
T_{\mu}T_{\nu}S^{\mu}S^{\nu}=T^{\lambda}T_{\rho}\,q_{\mu \nu \lambda}q^{\mu \nu \rho}=S^{\lambda}S_{\rho}\,q_{\mu \nu \lambda}q^{\mu \nu \rho}=T^{\lambda}S_{\rho}\,q_{\mu \nu \lambda}q^{\mu \nu \rho}=0\,,
\end{equation}

\begin{equation}
T_{\lambda}T_{\mu}T_{\nu}\,q^{\lambda \mu \nu}=S_{\lambda}S_{\mu}S_{\nu}\,q^{\lambda \mu \nu}=T_{\lambda}T_{\mu}S_{\nu}\,q^{\lambda \mu \nu}=T_{\lambda}S_{\mu}S_{\nu}\,q^{\lambda \mu \nu}=0\,,
\end{equation}

\begin{equation}
T_{\sigma}\,q_{\mu \nu \rho}\,q^{\mu \nu \lambda}\,q_{\lambda}\,^{\rho \sigma}=S_{\sigma}\,q_{\mu \nu \rho}\,q^{\mu \nu \lambda}\,q_{\lambda}\,^{\rho \sigma}=q_{\mu \nu \lambda}\,q^{\mu \nu \rho}\,q_{\sigma \omega}\,^{\lambda}\,q^{\sigma \omega}\,_{\rho}=q_{\lambda \sigma \mu}\,q^{\lambda \omega \nu}\,q^{\sigma}\,_{\rho \nu}\,q_{\omega}\,^{\rho \mu}=0\,.
\end{equation}

\section{Solutions}

By taking into account the previous remarks, the following constraints among the metric and torsion components are necessarily imposed together with the field equations and the basic space-time symmetry properties, in order to establish an appropriate physical consistency to the regarded PG model:

\begin{equation}
\Psi_{1} = \Psi_{2} \equiv \Psi(r)\,,
\end{equation}

\begin{equation}
b(r)=rc\,'(r)+c(r)+\frac{p}{r}\,,
\end{equation}

\begin{equation}
b(r) = a(r)\,\Psi(r)\,,\;\;\;c(r) = - \, g(r)\,\Psi(r)\,,\;\;\;d(r) = - \, h(r)\,\Psi(r)\,,\;\;\;l(r) = k(r)\,\Psi(r)\,.
\end{equation}

Note that these requirements do not demand the additional assumption of the double duality ansatz, usually considered by many authors due to its strong simplification of the field equations into a particular easier form \cite{Mielke}. Indeed, from a physical point of view, there is not any compelling reason to apply such a higher restriction, but a particular mathematical reduction in the difficulty of the computations, what in certain cases usually involves a loss of accuracy and generality that are incompatible with other possible configurations.

Then, the original model is appropriately simplified, and the following SO(3)-symmetric vacuum solution can be easily found for $c_{1} = -\,d_{1}/4\,, \, c_{2} = -\,d_{1}/2, \, \alpha = \frac{1}{2}\left(1-\beta\right)$ and $\gamma = -\,1$:

\begin{eqnarray}
a(r)&=&\frac{\Psi'(r)}{2\Psi(r)}+\frac{wr}{\Psi(r)}\,,\;\;
b(r)=\frac{\Psi'(r)}{2}+wr\,,\;\;
c(r)=\frac{\Psi(r)}{2r}+\frac{wr}{2}\,,\;\;
g(r)=-\,\frac{1}{2r}-\frac{wr}{2\Psi(r)}\,,\;\; \nonumber\\
d(r)&=&\frac{\kappa}{r}\,,\;\;\;\;\;\;\;\;\;\;\;\;\;\;\;\;\;\;\;\;\;
h(r)=-\,\frac{\kappa}{r\Psi(r)}\,,\;\;\;\;\;\;\,
k(r)=l(r)=0\;\;;
\end{eqnarray}
with

\begin{eqnarray}
\Psi(r)&=&1-\frac{2m}{r}+\frac{d_{1}\kappa^{2}}{r^2} \;\;,
\end{eqnarray}

\begin{equation}
w=\frac{\left(1-2\beta\right)}{d_{1}}\,.
\end{equation}

It is straightforward to note that the solution belongs to the special case where the contribution of the mass modes to the weak-field approximation of the torsion field is negligible. Then the relation (\ref{rel}) is completely fulfilled by taking $p=0$. In addition, the trace vector and the tensorial component of torsion remain massless whereas the axial mode becomes massless for $\beta=\frac{1}{2}$, what it means that our previous RN solution with massless torsion is recovered in such a case. This is an expected result, since it is shown that the dynamical behavior of torsion falls on the mentioned mode. Indeed, the axial component of torsion acts as a Coulomb-like potential depending on the parameter $\kappa$, which is related to the existence of a spin charge, in analogy to the relation between torsion and its spinning sources. Its geometrical effect is induced on the metric tensor by modifying the regular Schwarzschild vacuum structure of GR with the RN space-time associated with the following RC curvature tensor:

\begin{eqnarray}
F^{a b}\,_{c d} &=& \left(
\begin{array}{cccccc}
-\,w & 0 & 0 & 0 & 0 & 0 \\
0 & -\,w\,\chi_{-}(r)/2 & -\,\chi_{+}(r)\,(\kappa/2r^2) & 0 & -\,\chi_{+}(r)\,(\kappa/2r^2) & -w\,\chi_{+}(r)/2 \\
0 & \chi_{+}(r)\,(\kappa/2r^2) & -\,w\,\chi_{-}(r)/2 & 0 & w\,\chi_{+}(r)/2 & -\,\chi_{+}(r)\,(\kappa/2r^2) \\
-\,\kappa/r^2 & 0 & 0 & -\,\left(1/r^2+w/2\right) & 0 & 0 \\
0 & \chi_{-}(r)\,(\kappa/2r^2) & -\,w\,\chi_{+}(r)/2 & 0 & -3w\,\zeta(r)/2 & -\,\chi_{-}(r)\,(\kappa/2r^2) \\
0 & w\,\chi_{+}(r)/2 & \chi_{-}(r)\,(\kappa/2r^2) & 0 & \chi_{-}(r)\,(\kappa/2r^2) & -3w\,\zeta(r)/2
\end{array} \right)\,,
\end{eqnarray}
where the six rows and columns of the matrix are labeled the components in the order (01, 02, 03, 23, 31, 12) and the following functions have been defined:

\begin{equation}
\chi_{\pm}(r)=1\pm\frac{wr^{2}}{\Psi(r)}\,,
\end{equation}

\begin{equation}
\zeta(r)=1+\frac{wr^{2}}{3\Psi(r)}\,.
\end{equation}

Then, according to the first Bianchi identity in a RC space-time \cite{Ortin}, the solution reduces to the standard Schwarzschild geometry of GR when $\tilde{\nabla}_{[\mu}T^{\lambda}\,_{\nu \rho]}+T^{\sigma}\,_{[\mu \nu}\,T^{\lambda}\,_{\rho] \sigma}=0$, namely when the parameter $\kappa$ of the axial component vanishes.

It is also straightforward to notice the absence of singularities, excluding the point $r=0$, in the six independent quadratic scalar invariants defined from the curvature tensor, as expected from relations \eqref{rel2}:

\begin{equation}
\tilde{R}^{2}=\frac{4}{r^4}\left(1+6wr^2\right)^2\,,
\end{equation}

\begin{equation}
\tilde{R}_{\lambda \rho \mu \nu}\tilde{R}^{\lambda \rho \mu \nu}=\frac{4}{r^4}\left(1-\kappa^{2}+2wr^2\left(1+3wr^2\right)\right)\,,
\end{equation}

\begin{equation}
\tilde{R}_{\lambda \rho \mu \nu}\tilde{R}^{\mu \nu \lambda \rho}=\frac{4}{r^4}\left(1-2\kappa^{2}+2wr^2\left(1+3wr^2\right)\right)\,,
\end{equation}

\begin{equation}
\tilde{R}_{\lambda \rho \mu \nu}\tilde{R}^{\lambda \mu \rho \nu}=\frac{2}{r^4}\left(1-\kappa^{2}+2wr^2\left(1+3wr^2\right)\right)\,,
\end{equation}

\begin{equation}
\tilde{R}_{\mu \nu}\tilde{R}^{\mu \nu}=\frac{2}{r^4}\left(1+\kappa^{2}+6wr^2\left(1+3wr^2\right)\right)\,,
\end{equation}

\begin{equation}
\tilde{R}_{\mu \nu}\tilde{R}^{\nu \mu}=\frac{2}{r^4}\left(1-\kappa^{2}+6wr^2\left(1+3wr^2\right)\right)\,.
\end{equation}

On the other hand, the solution leads to a specific set of values for the Lagrangian coefficients, which should additionally define a viable and stable gravitational theory. According to the unitary and causality requirements, this consistency demands the absence of both ghosts and tachyons in the particle spectrum of the model, what has been systematically carried out by distinct approaches for the case of massive propagating torsion as well as for the case with zero-mass modes, where extra gauge symmetries can appear besides the fundamental Poincar\'e gauge symmetry \cite{Sezgin, Fukuma, Kuhfuss, Yo, Blagojevic, Battiti, Vasilev}. Nevertheless it should be noted that, apart from some particular differences and disagreements in their conclusions, all these approaches are not developed as perturbative analyses around any specific curved background which may be induced by the presence of a dynamical torsion, but on a rigid flat space-time where the possible effects of the torsion field are completely neglected. In fact, as can be seen, within our PG model the presence of a non-vanishing propagating torsion modifies the vacuum structure with the above RN geometry, where the axial component of the torsion tensor emerges in the metric tensor and hence it cannot be unilaterally excluded from the background. Furthermore, it is straightforward to note from \eqref{linear} that our PG model encompasses a weak-field approximation for the torsion field that cannot be separated from the background space-time (i.e. the torsion-free covariant derivatives of the Expression \eqref{linear} cannot be replaced by ordinary derivatives). Therefore, there exists a strong limitation around the cited stability studies, what it means that future analyses should be performed in order to examine the stability of these types of PG models.

Additionally, the solution can be naturally generalized to include the existence of a non-vanishing cosmological constant $\Lambda$ and Coulomb electromagnetic fields with electric and magnetic charges $q_{e}$ and $q_{m}$ respectively, which are decoupled from torsion under the assumption of the minimum coupling principle. This simple extension is obtained by modifying the metric function $\Psi(r)$ by the following expression:

\begin{equation}\Psi(r)=1-\frac{2m}{r}+\frac{d_{1}\kappa^{2}+q_{e}^{2}+q_{m}^{2}}{r^2}+\frac{\Lambda}{3}r^2\,.\end{equation}

Thereby, the solution shows similarities between the torsion and the electromagnetic fields, even though they are independent quantities. Note that it is referred to an extensive and regular PG theory, unlike other monopole-type solutions that can be constructed by modifying the model towards a different approach embedded within the complex Einstein-Yang-Mills theory \cite{Nakariki}.

The mass factors present in the solution may also involve corrections in the motion of spinning matter. Nevertheless, these deviations from the geodesic motion of ordinary matter are expected to be very small at astrophysics or cosmological scales, because of the vanishing of the spin density tensor in the most macroscopical bodies. This situation may differ around extreme gravitational systems as neutron stars or black holes with intense magnetic fields and sufficiently oriented elementary spins. In such a case, it is expected that the RC space-time described by the PG theory modulates these events. In addition, the influence of the mass of torsion on Dirac fields depends on the coupling considered between these and the torsion tensor. For Dirac fields minimally coupled to torsion, it turns out that only the axial vector carries out the interaction, whereas the trace vector and the tensorial mode are completely decoupled \cite{Hehl-Datta}. However, as can be seen from our RN solution, the parameter of mass associated with the axial mode falls on the rest of components of the torsion tensor, what it means that its effects may only be induced on Dirac fields non-minimally coupled to torsion.

\section{Conclusions}

In the present work, we have extended the correspondence between torsion and vacuum RN geometries in the framework of PG theory with massive torsion. This correspondence was first stressed in a previous work for the particular case given by a dynamical massless torsion alone, that can be associated with a PG model that contains quadratic order corrections in the curvature tensor \cite{CembGig}. Similar foundations were also introduced in \cite{Cembranos:2015eqa,Cembranos:2016kwh} in order to find an alternative method to solve the Einstein-Yang-Mills equations in extended gravitational theories. We investigate its generalization to the case with non-vanishing torsion mass modes by including the respective explicit torsion square pieces in the gravitational action. Then, we obtain the corresponding RN solution with massive torsion by imposing the appropriate space-time symmetries on the metric and torsion tensor, as well as additional consistency constraints in order to avoid all the possible unsuitable singularities and encompass the weak-field limit associated with torsion in a framework compatible with external Coulomb electric and magnetic fields, as in the standard case of GR.

In this scheme, the dynamical role of the torsion tensor is carried out by its axial mode, in a way that this mode can be massive or massless, whereas the mass modes of the trace vector and of the tensorial component remain vanishing. The presence of such a non-vanishing mass modifies the rest of the torsion components of the solution and it may introduce deviations in the trajectories of spinning matter. Nevertheless, it is shown that for the case of Dirac fields the non-minimal coupling to torsion is necessary. Even though, it is expected that the possible consequent effects are negligible at macroscopic scales and they may become significant only at extremely high densities.

Finally, the corresponding Reissner-Nordstr\"{o}m-de Sitter solution with cosmological constant and external electromagnetic fields is also obtained, by analogy with the standard case. The existence of these sorts of configurations reveals the dynamical role of the space-time torsion and provides new features associated with this field, what involves a richer vacuum structure of post-Riemannian gravitational theories endowed with both curvature and torsion.

\bigskip
\bigskip
\noindent
{\bf ACKNOWLEDGMENTS}

\bigskip

The authors acknowledge  F. W. Hehl for useful discussions. This work was partly supported by the projects FIS2014-52837-P (Spanish MINECO) and FIS2016-78859-P (AEI/FEDER, UE), and Consolider-Ingenio MULTIDARK CSD2009-00064.

\bigskip

\appendix
\section{Expressions of the field equations}

The Lagrangian (\ref{Lagrangian}) imposes the vanishing of the tensors $X1_{\mu}\,^{\nu}$ and $X2_{\mu}\,^{\lambda \nu}$ in vacuum, whose expressions can be written as:

\begin{equation}X1_{\mu}\,^{\nu} = -2\tilde{G}^{\nu}\,_{\mu}+4c_{1}T1_{\mu}\,^{\nu}+2c_{2}T2_{\mu}\,^{\nu}-2\left(2c_{1}+c_{2}\right)T3_{\mu}\,^{\nu}+2d_{1}\left(H1_{\mu}\,^{\nu}-H2_{\mu}\,^{\nu}\right)+\alpha \, I1_{\mu}\,^{\nu}+\beta \, I2_{\mu}\,^{\nu}+\gamma \, I3_{\mu}\,^{\nu}\,,\end{equation}

\begin{equation}\label{connectioneq}X2_{\mu}\,^{\lambda \nu} = \overset{\star}{T}_{\mu}\,^{\lambda \nu}+4c_{1}C1_{\mu}\,^{\lambda \nu}-2c_{2}C2_{\mu}\,^{\lambda \nu}+2\left(2c_{1}+c_{2}\right)C3_{\mu}\,^{\lambda \nu}-2d_{1}\left(Y1_{\mu}\,^{\lambda \nu}-Y2_{\mu}\,^{\lambda \nu}\right)-\alpha\,Z1_{\mu}\,^{\lambda \nu}-\beta\,Z2_{\mu}\,^{\lambda \nu}-\gamma\,Z3_{\mu}\,^{\lambda \nu}\,,\end{equation}
where it is given the explicit dependence with the following geometrical quantities:

\begin{equation}\tilde{G}_{\mu}\,^{\nu} = \tilde{R}_{\mu}\,^{\nu}-\frac{\tilde{R}}{2}\delta_{\mu}\,^{\nu}\,,\end{equation}

\begin{equation}T1_{\mu}\,^{\nu} = \tilde{R}_{\lambda \rho \mu \sigma}\tilde{R}^{\lambda \rho \nu \sigma}-\frac{1}{4}\delta_{\mu}\,^{\nu}\tilde{R}_{\lambda \rho \tau \sigma}\tilde{R}^{\lambda \rho \tau \sigma}\,,\end{equation}

\begin{equation}T2_{\mu}\,^{\nu} = \tilde{R}_{\lambda \rho \mu \sigma}\tilde{R}^{\lambda \nu \rho \sigma}+\tilde{R}_{\lambda \rho \sigma \mu}\tilde{R}^{\lambda \sigma \rho \nu}-\frac{1}{2}\delta_{\mu}\,^{\nu}\tilde{R}_{\lambda \rho \tau \sigma}\tilde{R}^{\lambda \tau \rho \sigma}\,,\end{equation}

\begin{equation}T3_{\mu}\,^{\nu} = \tilde{R}_{\lambda \rho \mu \sigma}\tilde{R}^{\nu \sigma \lambda \rho}-\frac{1}{4}\delta_{\mu}\,^{\nu}\tilde{R}_{\lambda \rho \tau \sigma}\tilde{R}^{\tau \sigma \lambda \rho}\,,\end{equation}

\begin{equation}H1_{\mu}\,^{\nu} = \tilde{R}^{\nu}\,_{\lambda \mu \rho}\tilde{R}^{\lambda \rho}+\tilde{R}_{\lambda \mu}\tilde{R}^{\lambda \nu}-\frac{1}{2}\delta_{\mu}\,^{\nu}\tilde{R}_{\lambda \rho}\tilde{R}^{\lambda \rho}\,,\end{equation}

\begin{equation}H2_{\mu}\,^{\nu} = \tilde{R}^{\nu}\,_{\lambda \mu \rho}\tilde{R}^{\rho \lambda}+\tilde{R}_{\lambda \mu}\tilde{R}^{\nu \lambda}-\frac{1}{2}\delta_{\mu}\,^{\nu}\tilde{R}_{\lambda \rho}\tilde{R}^{\rho \lambda}\,,\end{equation}

\begin{equation}I1_{\mu}\,^{\nu} = 4\left(T_{\lambda \rho \mu}T^{\lambda \rho \nu}+\nabla_{\lambda}T_{\mu}\,^{\nu \lambda}-K^{\rho}\,_{\mu \lambda}T_{\rho}\,^{\nu \lambda}-\frac{1}{4}\delta_{\mu}\,^{\nu}T_{\lambda \rho \sigma}T^{\lambda \rho \sigma}\right)\,,\end{equation}

\begin{equation}I2_{\mu}\,^{\nu} = 2\left(T_{\lambda \rho \mu}T^{\nu \rho \lambda}+T_{\lambda \rho \mu}T^{\rho \lambda \nu}+\nabla_{\lambda}T^{\lambda \nu}\,_{\mu}-\nabla_{\lambda}T^{\nu \lambda}\,_{\mu}+K^{\rho}\,_{\mu \lambda}\left(T^{\nu \lambda}\,_{\rho}-T^{\lambda \nu}\,_{\rho}\right)-\frac{1}{2}\delta_{\mu}\,^{\nu}T_{\lambda \rho \sigma}T^{\rho \lambda \sigma}\right)\,,\end{equation}

\begin{equation}I3_{\mu}\,^{\nu} = 2\left(T^{\nu}\,_{\mu \lambda}T^{\rho}\,_{\rho}\,^{\lambda}-\nabla_{\mu}T^{\lambda}\,_{\lambda}\,^{\nu}-K^{\nu}\,_{\mu \lambda}T^{\rho}\,_{\rho}\,^{\lambda}-\frac{1}{2}\delta_{\mu}\,^{\nu}\left(T^{\lambda}\,_{\lambda \sigma}T^{\rho}\,_{\rho}\,^{\sigma}-2\nabla_{\lambda}T^{\rho}\,_{\rho}\,^{\lambda}\right)\right)\,,\end{equation}

\begin{equation}\overset{\star}{T}_{\mu}\,^{\lambda \nu} = \delta_{\mu}\,^{\nu}g^{\lambda \sigma}T^{\rho}\,_{\rho \sigma}-g^{\lambda \nu}T^{\rho}\,_{\rho \mu}-g^{\lambda \sigma}T^{\nu}\,_{\mu \sigma}\,.\end{equation}

\begin{equation}C1_{\mu}\,^{\lambda \nu} = \nabla_{\rho}\tilde{R}_{\mu}\,^{\lambda \rho \nu}+K^{\lambda}\,_{\sigma \rho}\tilde{R}_{\mu}\,^{\sigma \rho \nu}-K^{\sigma}\,_{\mu \rho}\tilde{R}_{\sigma}\,^{\lambda \rho \nu}\,,\end{equation}

\begin{equation}C2_{\mu}\,^{\lambda \nu} = \nabla_{\rho}\left(\tilde{R}_{\mu}\,^{\nu \lambda \rho}-\tilde{R}_{\mu}\,^{\rho \lambda \nu}\right)+K^{\lambda}\,_{\sigma \rho}\left(\tilde{R}_{\mu}\,^{\nu \sigma \rho}-\tilde{R}_{\mu}\,^{\rho \sigma \nu}\right)-K^{\sigma}\,_{\mu \rho}\left(\tilde{R}_{\sigma}\,^{\nu \lambda \rho}-\tilde{R}_{\sigma}\,^{\rho \lambda \nu}\right)\,,\end{equation}

\begin{equation}C3_{\mu}\,^{\lambda \nu} = \nabla_{\rho}\tilde{R}^{\rho \nu \lambda}\,_{\mu}+K^{\lambda}\,_{\sigma \rho}\tilde{R}^{\rho \nu \sigma}\,_{\mu}-K^{\sigma}\,_{\mu \rho}\tilde{R}^{\rho \nu \lambda}\,_{\sigma}\,,\end{equation}

\begin{equation}Y1_{\mu}\,^{\lambda \nu}=\delta_{\mu}\,^{\nu}\nabla_{\rho}\tilde{R}^{\lambda \rho}-\nabla_{\mu}\tilde{R}^{\lambda \nu}+\delta_{\mu}\,^{\nu}K^{\lambda}\,_{\sigma \rho}\tilde{R}^{\sigma \rho}+K^{\rho}\,_{\mu \rho}\tilde{R}^{\lambda \nu}-K^{\nu}\,_{\mu \rho}\tilde{R}^{\lambda \rho}-K^{\lambda}\,_{\rho \mu}\tilde{R}^{\rho \nu}\,,\end{equation}

\begin{equation}Y2_{\mu}\,^{\lambda \nu} = \delta_{\mu}\,^{\nu}\nabla_{\rho}\tilde{R}^{\rho \lambda}-\nabla_{\mu}\tilde{R}^{\nu \lambda}+\delta_{\mu}\,^{\nu}K^{\lambda}\,_{\sigma \rho}\tilde{R}^{\rho \sigma}+K^{\rho}\,_{\mu \rho}\tilde{R}^{\nu \lambda}-K^{\nu}\,_{\mu \rho}\tilde{R}^{\rho \lambda}-K^{\lambda}\,_{\rho \mu}\tilde{R}^{\nu \rho}\,,\end{equation}

\begin{equation}Z1_{\mu}\,^{\lambda \nu} = 4T^{\lambda \nu}\,_{\mu}\,,\end{equation}

\begin{equation}Z2_{\mu}\,^{\lambda \nu} = 2\left(T^{\nu \lambda }\,_{\mu}-T^{\lambda \nu}\,_{\mu}\right)\,,\end{equation}

\begin{equation}Z3_{\mu}\,^{\lambda \nu} = g^{\lambda \nu}T^{\rho}\,_{\rho \mu}-\delta_{\mu}\,^{\nu}g^{\lambda \sigma}T^{\rho}\,_{\rho \sigma}\,.\end{equation}

\end{document}